\documentclass[conference]{IEEEtran}
\IEEEoverridecommandlockouts

\usepackage{cite}
\usepackage{amsmath,amssymb,amsfonts}
\usepackage{graphicx}
\usepackage{textcomp}
\usepackage{xcolor}
\def\BibTeX{{\rm B\kern-.05em{\sc i\kern-.025em b}\kern-.08em
		T\kern-.1667em\lower.7ex\hbox{E}\kern-.125emX}}
	
\usepackage{amsmath,amsfonts,amsthm,amssymb}
\usepackage{mathrsfs}
\usepackage{dutchcal}
\usepackage{algorithm}
\usepackage{algorithmicx}
\usepackage{algpseudocode}
\usepackage{epsfig}
\floatname{algorithm}{Algorithm}

\theoremstyle{remark}
\newtheorem{theorem}{\hskip 1em Theorem}

\newtheorem{property}{\hskip 1em Property}
\newtheorem{lemma}{\hskip 1em Lemma}

\newtheorem{assumption}{\hskip 1em Assumption}

\newtheorem{proof skecth}{Proof skecth}
\usepackage{array,caption}
\usepackage[mathscr]{eucal}
\usepackage[caption=false,font=LARGE,labelfont=rm,textfont=rm]{subfig}
\usepackage{textcomp}
\usepackage{stfloats}
\usepackage{url}
\usepackage{verbatim}
\usepackage{graphicx,float,balance}
\usepackage{multirow}
\usepackage{cite}
\usepackage{soul}
\usepackage{color,xcolor}
\begin{document}	
	\title{Perturbation Power Selection for First-Error Delay Maximization in Enhanced SC Decoding
	}
	
	\author{
		\IEEEauthorblockN{Zhicheng Liu\IEEEauthorrefmark{1}, Liuquan Yao\IEEEauthorrefmark{2}, Shuai Yuan\IEEEauthorrefmark{2}, Zhongjun Yang\IEEEauthorrefmark{3}, Guiying Yan\IEEEauthorrefmark{2},\\ Zhiming Ma\IEEEauthorrefmark{2}, Li Chen\IEEEauthorrefmark{3}, and Zechun Hu\IEEEauthorrefmark{1}}
		\IEEEauthorblockA{\IEEEauthorrefmark{1}%
			School of Mathematics, Sichuan University, Chengdu, China} 
		\IEEEauthorblockA{\IEEEauthorrefmark{2}%
			Academy of Mathematics and Systems Science, University of Chinese Academy and Sciences, Beijing, China}
		\IEEEauthorblockA{\IEEEauthorrefmark{3}%
			School of Electronics and Information Technology, Sun Yat-sen University, Guangzhou, China}
		Email: lzc2025@scu.edu.cn, yaoliuquan20@mails.ucas.ac.cn, yuanshuai2020@amss.ac.cn, yangzhj59@mail2.sysu.edu.cn, \\ 
		yangy@amss.ac.cn, mazm@amt.ac.cn, chenli55@mail.sysu.edu.cn, zchu@scu.edu.cn
	}

\maketitle

\begin{abstract}
	In this paper, we analyze the effect of perturbation power in delaying the first error position (FEP), i.e., the first information bit incorrectly decoded by the successive cancellation (SC) decoding. It is conducted over the finite-length perturbation-enhanced SC (PE-SC) decoding paradigm. We show that the FEP delaying probability exhibits a non-monotonic dependence on the perturbation power. Based on this property, an efficient perturbation power selection algorithm that maximizes the delay probability is proposed to enhance the perturbation efficiency. It results in a more efficient perturbation power selection in finite-length PE-SC decoding.
\end{abstract}


\section{Introduction}
\IEEEPARstart{P}{olar} codes, introduced by Arıkan~\cite{ref1}, achieve capacity-approaching performance and have been adopted for 5G eMBB control channels~\cite{ref2}, motivating extensive research on advanced decoding algorithms. Although SC decoding has low complexity, its finite-length performance remains limited~\cite{ref1}. Addressing this issue, SC list (SCL) decoding~\cite{ref3,ref4} and cyclic redundancy check (CRC) aided SCL (CA-SCL) decoding~\cite{ref5,adaptive_SCL_CRC} have been proposed. These algorithms can significantly improve the SC decoding performance, but at the cost of huge computational complexity.

Further improvements have been explored via flipping-based decoding~\cite{ref6,ref7,T_SCF_Ercan,Threshold_SCF} and automorphism ensemble (AE) decoding~\cite{ref8,ref9,ref10,Group_AE_polar}. However, their effectiveness deteriorates for long polar codes due to the growing set of unreliable bits~\cite{ref11} and the asymptotically vanishing useful automorphism groups in Arıkan-constructed polar codes~\cite{ref12}. 

To address these challenges, perturbation-enhanced (PE) decoding has emerged as a scalable alternative for long polar codes~\cite{ref13,ref14,ref16}, where random perturbations are injected into the log-likelihood ratios (LLRs) when the CRC detection fails. Recent studies further improved PE-SC/SCL decoding by exploiting historical information~\cite{adaptive_PSCL} and identifying unreliable information positions~\cite{Yang_Improved_SC_perturb}. Nevertheless, perturbation design in existing PE decoding schemes remains largely empirical due to the lack of rigorous theoretical analysis. Motivated by the asymptotic non-degrading property established in~\cite{ref16}, we focus on the finite-length behavior of PE-SC decoding. In particular, we investigate the effect of perturbation power on delaying the first error position (FEP) and develop a perturbation power selection algorithm to enhance the perturbation efficiency. The proposed criterion can be further generalized into adaptive perturbation frameworks.

The main contributions are summarized as follows:
\begin{enumerate}
	\item We show that, for a given codeword length $N$, the FEP delaying probability $\mathbb{P}(\mathrm{delay})$ exhibits a non-monotonic dependence on the perturbation power $\sigma_p^2$.
	\item We reveal that properly calibrated perturbation improves PE-SC decoding performance: insufficient perturbation yields limited deviation from the original signal, whereas excessive perturbation causes the perturbed signal to be dominated by perturbation noise.
	\item We develop a perturbation power selection algorithm that maximizes $\mathbb{P}(\mathrm{delay})$ in the finite-length regime.
\end{enumerate}

\textit{Notations Convention:} 
The probability measure and expectation are represented by $\mathbb{P}(\cdot)$ and $\mathbb{E}(\cdot)$, respectively. The complementary cumulative distribution function of the standard normal distribution is defined as $\mathrm{Q}(x) = \frac{1}{\sqrt{2\pi}} \int_x^{+\infty} e^{-\frac{t^2}{2}} \mathrm{d}t$. Boldface letters denote vectors and matrices. The notation $\mathrm{o}(h)$ denotes a higher-order infinitesimal term that vanishes faster than $h$ as $h \to 0$ and $\mathrm{sgn}(\cdot)$ is the sign function.

\section{Preliminaries}
\subsection{Polar Codes}
A polar code of length $N=2^n$ is defined by
\begin{flalign*}
	\mathbf{v}_{0:N} = \mathbf{u}_{0:N} \mathbf{G}_{N}, 
\end{flalign*}
where \(\mathbf{v}_{0:N} = [v_{0}, \ldots, v_{N-1}]\) denotes the codeword, \(\mathbf{u}_{0:N}=[u_{0}, \ldots, u_{N-1}]\) represents the source bit, and \(\mathbf{G}_{N}=\mathbf{F}^{\otimes n}\) with \(\mathbf{F} = \begin{bmatrix}\begin{smallmatrix}
		1 & 0 \\ 1 & 1
\end{smallmatrix}  \end{bmatrix}\), and $\mathbf{F}^{\otimes n}$ is the
$n$-fold Kronecker product of $\mathbf{F}$. The information set
$\mathscr{A}$ contains the $K$ most reliable polarized bit-channels,
while the remaining positions are frozen to zero.

We consider binary phase-shift keying (BPSK) transmission over an
additive white Gaussian noise (AWGN) channel. Each coded bit $v_i$ is
mapped to $s_i=1-2v_i$, yielding
\[
y_i=s_i+n_i,\qquad n_i\sim\mathscr{N}(0,\sigma^2),
\]
where $\sigma^2$ denotes the channel-noise variance (power). The
corresponding channel LLR vector is
$\mathbf{L}_{0:N}:=[L_0,\ldots,L_{N-1}]
=2\mathbf{y}/\sigma^2$, where
$\mathbf{y}:=[y_0,\ldots,y_{N-1}]$.
%
%
	
\subsection{SC Decoding}
Let the received LLR corresponding to $y_i$ be defined as $L_1^{(i)}=2y_i/\sigma^2$. For SC decoding, let $L_N^{(i)}$ (or simply $L_i$) denote the decision LLR of bit $u_i$ \cite[Eq.~(7)]{ref4}, given by
\begin{flalign*}
	L_{N}^{(i)}
	=
	\ln
	\frac{
		W_{N}^{(i)}(\mathbf{y}_{0:N},\hat{\mathbf{u}}_{0:i}\mid u_i=0)
	}{
		W_{N}^{(i)}(\mathbf{y}_{0:N},\hat{\mathbf{u}}_{0:i}\mid u_i=1)
	},
\end{flalign*}
where $W_{N}^{(i)}(\cdot)$ denotes the $i$-th synthesized subchannel \cite{ref1}. The SC decision rule is
\begin{flalign*}
	\hat{u}_{i}=
	\begin{cases}
		0, & L_{N}^{(i)}>0,\ i\in\mathscr{A},\\
		1, & L_{N}^{(i)}<0,\ i\in\mathscr{A},\\
		u_i, & i\in\mathscr{A}^{c}.
	\end{cases}
\end{flalign*}

For a length-$2$ polar code with $\mathbf{u}_{0:1}=[u_0,u_1]$, the decision LLR of $u_0$ is first computed via the min-sum approximation of the $f$-function~\cite[Eq.~(10)]{MSA_f_function_SC},
\begin{flalign*}
	L_{2}^{(0)}
	=
	f(L_1^{(0)},L_1^{(1)})
	=
	\mathrm{sgn}(L_1^{(0)}L_1^{(1)})
	\min\{|L_1^{(0)}|,|L_1^{(1)}|\}.
\end{flalign*}
After obtaining $\hat{u}_0$, the decision LLR of $u_1$ is updated through the $g$-function as
\begin{flalign*}
	L_{2}^{(1)}
	=
	g(L_1^{(0)},L_1^{(1)},\hat{u}_0)
	=
	(1-2\hat{u}_0)L_1^{(0)}+L_1^{(1)}.
\end{flalign*}
A hard decision on $L_{2}^{(1)}$ then yields $\hat{u}_1$.

For a length-$N=2^n$ polar code, SC decoding recursively applies the $f$- and $g$-functions to generate the estimate
$\hat{\mathbf{u}}_{0:N}=[\hat{u}_0,\ldots,\hat{u}_{N-1}]$, where $\hat{u}_j=u_j$ for all $j\in\mathscr{A}^{c}$.

\begin{algorithm}[htbp]
	\caption{SC specialization of PE decoding~\cite{ref14}}
	\label{alg:PE-SC}
	\begin{algorithmic}[1]
		\Require Channel LLR vector $\mathbf{L}_{0:N}$, perturbation variance
		$\sigma_p^2$, total number of attempts $T$, and information set
		$\mathscr{A}$
		\Ensure Decoded vector $\hat{\mathbf{u}}$ or decoding failure
		
		\State $\hat{\mathbf{u}}_{0:N}
		\gets \operatorname{SCDecoder}(\mathbf{L}_{0:N},\mathscr{A})$
		\If{CRCCheck$(\hat{\mathbf{u}}_{0:N})=\text{fail}$}
		\For{$t=1$ to $T-1$ }
		\State Draw $\mathbf{n}^{(t)}
		\sim\mathscr{N}(\mathbf{0},\sigma_p^2\mathbf{I}_N)$
		\State $\hat{\mathbf{u}}_{0:N}
		\gets \operatorname{SCDecoder}
		(\mathbf{L}_{0:N}+\mathbf{n}^{(t)},\mathscr{A})$;
		\If{$\operatorname{CRCCheck}(\hat{\mathbf{u}})
			=\mathrm{success}$}
		\Return $\hat{\mathbf{u}}_{0:N}$;
		\EndIf
		\EndFor
		\EndIf
		\State \Return decoding failure
	\end{algorithmic}
\end{algorithm}
\subsection{PE-SC Decoding}
We consider the SC specialization of the perturbation-enhanced decoding principle originally proposed for SCL decoding in~\cite{ref14}, as summarized in Algorithm~\ref{alg:PE-SC}. Upon CRC failure, the decoder performs at most $T-1$ perturbed SC attempts using $\mathbf{L}_{0:N}+\mathbf{n}^{(t)}$, where $\mathbf{n}^{(t)}\sim\mathscr{N} (\mathbf{0},\sigma_p^2\mathbf{I}_N)$ is independently sampled across coordinates and attempts, and $\mathbf{I}_N$ is the $N\times N$ identity matrix. Here, $T$ is the total number of decoding attempts; thus, $T=2$ represents one conventional SC attempt followed by one perturbed attempt, and $\sigma_p^2$ denotes the perturbation power. Conceptually, perturbation explores nearby LLR realizations, potentially correcting an erroneous early decision while retaining useful channel information; excessive perturbation may instead introduce new errors.

\subsection{FEP Delaying Probability}
Denote $\hat{\mathbf{u}}^{(0)}=[\hat{u}^{(0)}_{0},\ldots,\hat{u}^{(0)}_{N-1}]$, $\hat{\mathbf{u}}^{(1)}=[\hat{u}^{(1)}_{0},\ldots,\hat{u}^{(1)}_{N-1}]$ the decoding outputs of the SC decoder and the PE-SC decoder with a single perturbation. Let $\tau_0$ and $\tau_1$ denote the indices of the FEP over the above two SC decoders, respectively, i.e., the first bit $i\in\mathscr{A}$ yielding $\hat{u}^{(0)}_{i}\neq u_i$ and $\hat{u}^{(1)}_{i}\neq u_i$. We then define the following conditional probabilities:
\begin{flalign*}
	&\mathbb{P}(\mathrm{delay}) 
	\triangleq \mathbb{P}\!\left(\tau_1 > \tau_0 \mid \tau_0 < N\right),\\
	&\mathbb{P}(\mathrm{unchanged}) 
	\triangleq \mathbb{P}\!\left(\tau_1 = \tau_0 \mid \tau_0 < N\right),\\
	&\mathbb{P}(\mathrm{advance}) 
	\triangleq \mathbb{P}\!\left(\tau_1 < \tau_0 \mid \tau_0 < N\right).
\end{flalign*}

It was shown in~\cite[Theorem~1]{ref16} that both the FEP delaying probability $\mathbb{P}(\mathrm{delay})$ and the FEP unchanged probability $\mathbb{P}(\mathrm{unchanged})$ converge to $\frac{1}{2}$ asymptotically. However, it is not yet understood how $\mathbb{P}(\mathrm{delay})$ behaves when the asymptoticity vanishies, i.e., over the finite codeword length regime.

\section{Impact of Perturbation Power on FEP Delaying Probability}
In this section, we study the dependence of $\mathbb{P}(\mathrm{delay})$ on the perturbation parameter $\sigma_p$. We show that $\mathbb{P}(\mathrm{delay})$ exhibits a non-monotonic behavior with respect to $\sigma_p$. It implies that in maxmizing $\mathbb{P}(\mathrm{delay})$, an optimal perturbation parameter $\sigma_{p}$ should be chosen. This property motivates a perturbation power selection criterion that maximizes $\mathbb{P}(\mathrm{delay})$ for a given code length $N$ and information set $\mathscr{A}$.

\subsection{Dependence of $\mathbb{P}(\mathrm{delay})$ on Perturbation Power $\sigma_{p}^{2}$}
In this subsection, we analyze the dependence of $\mathbb{P}(\mathrm{delay})$ on $\sigma_p$. Without loss of generality, we assume transmission of the all-zero codeword $\mathbf{0}_N$ over a symmetric channel~\cite{ref15}, and let $\mathbf{0}$ denote an all-zero vector of an arbitrary length. Let $\mu_N^{(i)}$ denote the mean of $L_{N}^{(i)}$. For the underlying channel, $\mu_1^{(i)} = \frac{2}{\sigma^2}$ for all $i \in \{0,\ldots,N-1\}$ \cite[Section~III]{ref15}.

\begin{assumption}[Gaussian Approximation (GA)~\cite{ref15}]\label{asm1}
	Let us assume each subchannel LLR follows a constrained Gaussian distribution whose variance is twice its mean, and conditioned on $\hat{\mathbf{u}}^{(0)}_{0:i}=\mathbf{0}$ when decoding $u_i$, we can obtain
	\begin{flalign*}
		\{L_{i}\mid \hat{\mathbf{u}}^{(0)}_{0:i}=\mathbf{0}\} \sim \mathscr{N}(\mu_{i}, 2\mu_{i}),
	\end{flalign*} 
	where $\mu_N^{(i)}$ (abbreviated as $\mu_{i}$) is recursively given as follows:
	\begin{flalign*}
		\mu^{(i)}_{N}=
		\begin{cases}
			\phi^{-1}(1-(1-\phi(\mu^{(\frac{i}{2})}_{\frac{N}{2}}))^{2}), &\text{if}\ i\in \{0,2,\cdots,N-2\}, \\
			2 \mu_{\frac{N}{2}}^{(\frac{i-1}{2})}, &\text{if}\ i\in\{1,3,\cdots,N-1\},
		\end{cases}
	\end{flalign*}
	and
	\begin{flalign*}
		\phi(x)\triangleq 1-\int_{-\infty}^{\infty}\frac{1}{\sqrt{4\pi x}}\tanh(\frac{t}{2})e^{-\frac{(t-x)^{2}}{4x}} \mathrm{d}t.
	\end{flalign*}
\end{assumption}

The following property from~\cite[Proposition~1]{ref16} provides a generalized GA characterization for PE-SC decoding and is used repeatedly in the finite-length analysis of $\mathbb{P}(\mathrm{delay})$.
\begin{property}\label{asm2}
	For all information bits with $i \in \mathscr{A}$,
	\begin{flalign*}
		\{L_i, L_i + n_{u_i} \mid \hat{\mathbf{u}}^{(0)}_{0:i}=\mathbf{0}, \hat{\mathbf{u}}^{(1)}_{0:i}=\mathbf{0}\}\sim (G_1, G_1 + G_2),
	\end{flalign*}
	where $n_{u_i}\sim\mathscr{N}(0,\sigma_i^2)$, $G_1\sim\mathscr{N}(\mu_i,2\mu_i)$ and $G_2\sim\mathscr{N}(0,\sigma_i^2)$ are independent, where $\sigma_i^2=2^{k_i}\sigma_p^2$ and $k_i$ is the number of $g$-functions along the SC decoding path leading to bit $u_i$ \cite{ref16}.
\end{property}
Next, we will present several important results that form the basis for the proof of major conclusion.
\begin{property}\cite[Eq.~(9)]{ref17} \label{prop2}
	For any $ x > 0 $, $ \mathrm{Q}(x) \approx \frac{1}{2} \exp{(-\frac{x^{2}}{2})} $.
\end{property}
\begin{property} \label{prop3}
	Given that $ a, b, c > 0 $, we observe:
	\begin{flalign*}
		&\sqrt{\frac{ax}{ax+b}} \approx \sqrt{\frac{a}{b}} x^{\frac{1}{2}} - \frac{1}{2} \left( \frac{a}{b} \right)^{\frac{3}{2}} x^{\frac{3}{2}} + \frac{3}{8} \left( \frac{a}{b} \right)^{\frac{5}{2}} x^{\frac{5}{2}} + \mathrm{o}(x^{\frac{5}{2}}), \\
		&e^{-\frac{c}{a+bx}} \approx e^{-\frac{c}{a}} + \frac{bc e^{-\frac{c}{a}}}{a^2} x - \frac{b^2 (2a - c) c e^{-\frac{c}{a}}}{2a^4} x^2 + \mathrm{o}(x^2).
	\end{flalign*}
\end{property}
\begin{proof}
	Examining Taylor series expansions of these functions around $ x = 0 $, further simplifications yield the proof.
\end{proof}
The lemma below characterizes the probability that a single perturbation corrects an initially misdecoded bit $u_{i}$, conditioned on correct decoding of all preceding bits.
\begin{lemma} \label{lem1}
	Let $L_{i}\sim\mathscr{N}(\mu_{i},2\mu_{i})$ and $n_{u_{i}}\sim\mathscr{N}(0,\sigma_i^2)$ be independent, we obtain
	\begin{flalign}
		\mathbb{P}(L_{i} < 0, L_{i} + n_{u_{i}} > 0) \approx \frac{1}{4} \sqrt{\frac{\sigma_i^2}{\sigma_i^2 + 2\mu_{i}}} e^{-\frac{\mu_{i}}{4}}.
	\end{flalign}
\end{lemma}
\begin{proof}
Noting that $L$ and $n_{u}$ are independent, we get
\begin{flalign*}
	&\mathbb{P}(L_{i}<0,L_{i}+n_{u_{i}}>0)=\int_{-\infty}^{-\sqrt{\frac{\mu_{i}}{2}}}\frac{e^{-\frac{x^{2}}{2}}\mathrm{Q}(-\frac{\sqrt{2\mu_{i}}x+\mu_{i}}{\sigma_{i}})}{\sqrt{2\pi}}\mathrm{d}x\\
	&\overset{\frac{\sqrt{2\mu_{i}}x+\mu_{i}}{\sigma_{i}}=t}{=\!=\!=\!=}\int_{-\infty}^{0}\frac{e^{-\frac{(\sigma_{i}t-\mu_{i})^{2}}{4\mu_{i}}}}{\sqrt{2\pi}}\mathrm{Q}(-t)\frac{\sigma_{i}}{\sqrt{2\mu_{i}}}\mathrm{d}t\\
	&\approx \frac{\sigma_{i}}{2\sqrt{2\mu_{i}}} \int_{-\infty}^{0}\frac{1}{\sqrt{2\pi}}e^{-\frac{(\sigma_{i}t-\mu_{i})^{2}}{4\mu_{i}}}e^{-\frac{t^{2}}{2}}\mathrm{d}t.
\end{flalign*}

By transforming $\frac{t-\frac{\sigma_{i}\mu_{i}}{\sigma_{i}^{2}+2\mu_{i}}}{\sqrt{\frac{2\mu_{i}}{\sigma_{i}^{2}+2\mu_{i}}}}=s$, we observe
\begin{flalign*}
	&\mathbb{P}(L_{i}<0,L_{i}+n_{u_{i}}>0)\\
	\approx&\frac{e^{-\frac{\mu_{i}^{2}}{2(\sigma_{i}^{2}+2\mu_{i})}}}{2}\sqrt{\frac{\sigma_{i}^{2}}{\sigma_{i}^{2}+2\mu_{i}}}\int_{-\infty}^{-\sqrt{\frac{\mu_{i}\sigma_{i}^{2}}{2(\sigma_{i}^{2}+2\mu_{i})}}}\frac{1}{\sqrt{2\pi}}e^{-\frac{s^{2}}{2}}\mathrm{d}s\\
	=&\frac{1}{2}\sqrt{\frac{\sigma_{i}^{2}}{\sigma_{i}^{2}+2\mu_{i}}}e^{-\frac{\mu_{i}^{2}}{2(\sigma_{i}^{2}+2\mu_{i})}}\mathrm{Q}(\sqrt{\frac{\mu_{i}\sigma_{i}^{2}}{2(\sigma_{i}^{2}+2\mu_{i})}})\\
	\approx&\frac{1}{4}\sqrt{\frac{\sigma_{i}^{2}}{\sigma_{i}^{2}+2\mu_{i}}}e^{-\frac{\mu_{i}^{2}}{2(\sigma_{i}^{2}+2\mu_{i})}}e^{-\frac{\mu_{i}\sigma_{i}^{2}}{4(\sigma_{i}^{2}+2\mu_{i})}}
	=\frac{e^{-\frac{\mu_{i}}{4}}}{4}\sqrt{\frac{\sigma_{i}^{2}}{\sigma_{i}^{2}+2\mu_{i}}}.
\end{flalign*}

Therefore, the conclusion can be reached.
\end{proof}
The following lemma quantifies the probability that an initially correctly decoded bit $u_{i}$ remains correct after perturbation, where all preceding bits have been decoded correctly.
	\begin{lemma}\label{lem2}
		With $L_{i}\sim\mathscr{N}(\mu_{i},2\mu_{i})$ and $n_{u_{i}}\sim\mathscr{N}(0,\sigma_i^2)$ independent, it follows that
		\begin{flalign}
			&\mathbb{P}(L_{i}>0,L_{i}+n_{u_{i}}>0)\approx\mathrm{Q}(-\sqrt{\frac{\mu_{i}}{2}})+\frac{1}{4}\sqrt{\frac{\sigma_{i}^{2}}{\sigma_{i}^{2}+2\mu_{i}}}e^{-\frac{\mu_{i}}{4}}\nonumber\\
			&-\frac{1}{2}\sqrt{\frac{\sigma_{i}^{2}}{\sigma_{i}^{2}+2\mu_{i}}}e^{-\frac{\mu_{i}^{2}}{2(\sigma_{i}^{2}+2\mu_{i})}}.
		\end{flalign}
	\end{lemma}
	\begin{proof}
		Using steps similar to those in Lemma \ref{lem1}, we have
		\begin{flalign*}
			&\mathbb{P}(L_{i}>0,L_{i}+n_{u_{i}}>0)
			=\int_{-\sqrt{\frac{\mu_{i}}{2}}}^{+\infty}\frac{e^{-\frac{x^{2}}{2}}\mathrm{Q}(-\frac{\sqrt{2\mu_{i}}x+\mu_{i}}{\sigma_{i}})}{\sqrt{2\pi}}\mathrm{d}x\\
			&=\mathrm{Q}(-\sqrt{\frac{\mu_{i}}{2}})-\int_{-\sqrt{\frac{\mu_{i}}{2}}}^{+\infty}\frac{1}{\sqrt{2\pi}}e^{-\frac{x^{2}}{2}}\mathrm{Q}(\frac{\sqrt{2\mu_{i}}x+\mu_{i}}{\sigma_{i}})\mathrm{d}x\\
			&\overset{\frac{\sqrt{2\mu_{i}}x+\mu_{i}}{\sigma_{i}}=t}{=\!=\!=\!=\!=}\mathrm{Q}(-\sqrt{\frac{\mu_{i}}{2}})-\frac{\sigma_{i}}{2\sqrt{\pi\mu_{i}}}\int_{0}^{+\infty}e^{-\frac{(\sigma_{i}t-\mu_{i})^{2}}{4\mu_{i}}}\mathrm{Q}(t)\mathrm{d}t\\
			&\approx \mathrm{Q}(-\sqrt{\frac{\mu_{i}}{2}})-\frac{\sigma_{i}}{2\sqrt{2\mu_{i}}}\int_{0}^{+\infty}\frac{1}{\sqrt{2\pi}}e^{-\frac{(\sigma_{i}t-\mu_{i})^{2}}{4\mu_{i}}}e^{-\frac{t^{2}}{2}}\mathrm{d}t.
		\end{flalign*}
		
		Letting $\frac{t - \frac{\sigma_i \mu_{i}}{\sigma_i^2 + 2\mu_{i}}}{\sqrt{\frac{2\mu_{i}}{\sigma_i^2 + 2\mu_{i}}}} = s$, additional calculations arrive at
		\begin{flalign*}
			&\mathbb{P}(L_{i} > 0, L_{i} + n_{u_{i}} > 0) \\
			\approx& \mathrm{Q}(-\sqrt{\frac{\mu_{i}}{2}}) -\frac{1}{2} \sqrt{\frac{\sigma_i^2}{\sigma_i^2 + 2\mu_{i}}} e^{-\frac{\mu_{i}^2}{2(\sigma_i^2 + 2\mu_{i})}} \mathrm{Q}(-\sqrt{\frac{\mu_{i} \sigma_i^2}{2(\sigma_i^2 + 2\mu_{i})}})\\
			\approx& \mathrm{Q}(-\sqrt{\frac{\mu_{i}}{2}}) -\frac{1}{2} \sqrt{\frac{\sigma_i^2}{\sigma_i^2 + 2\mu_{i}}} e^{-\frac{\mu_{i}^2}{2(\sigma_i^2 + 2\mu_{i})}} [1 -\frac{1}{2} e^{-\frac{\mu_{i}\sigma_{i}^{2}}{4(\sigma_{i}^{2}+2\mu_{i})}}]\\
			\approx& \mathrm{Q}(-\sqrt{\frac{\mu_{i}}{2}}) -\frac{1}{2} \sqrt{\frac{\sigma_i^2}{\sigma_i^2 + 2\mu_{i}}} \left[e^{-\frac{\mu_{i}^2}{2(\sigma_i^2 + 2\mu_{i})}} - \frac{1}{2}e^{-\frac{\mu_{i}}{4}}\right],
		\end{flalign*}
		which leads us to the conclusion of Lemma \ref{lem2}.
	\end{proof}
Denote $D_N(\sigma_p):=\Pr(\tau_1>\tau_0
\mid\tau_0<N)$ (or simply $D_N$) the FEP delaying probability $\mathbb{P}(\mathrm{delay})$ conditioned on
failure of the initial SC attempt. The following theorem provides a
third-order approximation to $D_N(\sigma_p)$ and yields a
low-complexity rule for selecting the perturbation variance
$\sigma_p^2$.
Let
\begin{equation*}
	\left\{
	\begin{aligned}
		A_i &= a_i b_i \prod_{j \in \mathscr{A},\, j < i} c_j, \\
		B_i &= \tfrac{1}{4} a_i b_i \prod_{j \in \mathscr{A},\, j < i} a_j b_j, \\
		C_i &= \tfrac{1}{2} a_i b_i^3 \prod_{j \in \mathscr{A},\, j < i} c_j,
	\end{aligned}
	\right.
\end{equation*}
where $a_i=e^{-\mu_i/4}$, $b_i=\sqrt{2^{k_i-1}/\mu_i}$, and $c_i=\mathrm{Q}(-\sqrt{\mu_i/2})$ for all $i\in\mathscr{A}$. Here, $\mu_i$ denotes the mean decision LLR of $u_i$, and $k_i$ denotes the number of $g$-functions encountered along the SC decoding path up to $u_i$, equivalently given by the Hamming weight of the binary representation of $i-1$~\cite{ref16}.

\begin{theorem}\label{thm1}
	For a fixed code length $N$ and information set $\mathscr{A}$, the
	FEP delaying probability admits the approximation
	\begin{equation}
		D_N(\sigma_p)
		\approx
		\frac{A_{\Sigma}\sigma_p-B_{\Sigma}\sigma_p^2-C_{\Sigma}\sigma_p^3}
		{4\mathrm{BLER}_{\mathrm{SC}}},
		\label{eq10}
	\end{equation}
	where $A_{\Sigma}=\sum_{i\in\mathscr{A}}A_i$,
	$B_{\Sigma}=\sum_{i\in\mathscr{A}}B_i,$ and
	$C_{\Sigma}=\sum_{i\in\mathscr{A}}C_i$.
	Differentiating the right-hand side of~\eqref{eq10} shows that, within
	the validity range of the approximation, it increases up to
	\begin{equation}
		\sigma_p^*
		=
		\frac{
			\sqrt{B_{\Sigma}^{2}+3A_{\Sigma}C_{\Sigma}}-B_{\Sigma}
		}{
			3C_{\Sigma}
		}.
		\label{eq11}
	\end{equation}
	and decreases thereafter. Thus, $\sigma_p^*$ maximizes the third-order approximation and provides
	an approximately optimal perturbation standard deviation.
\end{theorem}
\begin{proof}
	Recalling the definition of $\mathbb{P}(\mathrm{delay})$, we deduce
	\begin{flalign*}
		&D_N(\sigma_p)=\mathbb{P}(\tau_1>\tau_0\mid\tau_0<N)\\
		&=\frac{\mathbb{P}(\tau_1>\tau_0,\tau_0<N)}{\mathbb{P}(\tau_0<N)}=\frac{\sum_{i\in\mathscr{A}}\mathbb{P}(\tau_1>i,\tau_0=i)}{\sum_{i\in\mathscr{A}}\mathbb{P}(\hat{u}^{(0)}_{i}=1,\hat{\mathbf{u}}^{(0)}_{0:i}=\mathbf{0})}\\
		&=\frac{\sum_{i\in\mathscr{A}}\mathbb{P}(\tau_1>i,\tau_0=i)}{\mathbb{P}(\exists i\in\mathscr{A}, \mathrm{s.t.,} \hat{u}^{(0)}_{i}=1)}= \frac{\sum_{i\in\mathscr{A}}\mathbb{P}(\tau_1>i,\tau_0=i)}{\mathrm{BLER_{SC}}}.
	\end{flalign*}
	
	Using the chain rule of conditional probability, we obtain:
	\begin{flalign*}
		&\mathbb{P}(\tau_1>i,\tau_0=i)\\
		&=\mathbb{P}(\hat{u}^{(1)}_{i}=0,\hat{u}^{(0)}_{i}=1,\hat{\mathbf{u}}^{(1)}_{0:i}=\mathbf{0},\hat{\mathbf{u}}^{(0)}_{0:i}=\mathbf{0})\\
		&=\mathbb{P}(\hat{u}^{(1)}_{i}=0,\hat{u}^{(0)}_{i}=1\mid\hat{\mathbf{u}}^{(1)}_{0:i}=\mathbf{0},\hat{\mathbf{u}}^{(0)}_{0:i}=\mathbf{0})\\
		&\prod_{j<i,j\in\mathscr{A}}\mathbb{P}(\hat{u}^{(1)}_{j}=0,\hat{u}^{(0)}_{j}=0\mid\hat{\mathbf{u}}^{(1)}_{0:j}=\mathbf{0},\hat{\mathbf{u}}^{(0)}_{0:j}=\mathbf{0})\\
		&=\mathbb{P}(L_{i}+n_{u_{i}}>0,L_{i}<0\mid\hat{\mathbf{u}}^{(1)}_{0:i}=\mathbf{0},\hat{\mathbf{u}}^{(0)}_{0:i}=\mathbf{0})\\
		&\prod_{j<i,j\in\mathscr{A}}\mathbb{P}(L_{j}+n_{u_{j}}>0,L_{j}>0\mid\hat{\mathbf{u}}^{(1)}_{0:j}=\mathbf{0},\hat{\mathbf{u}}^{(0)}_{0:j}=\mathbf{0}).
	\end{flalign*}
	where $L_i \sim \mathscr{N}(\mu_i, 2\mu_i)$ and $n_{u_i}\sim\mathscr{N}(0,\sigma_i^2)$.
	
	By applying Lemma \ref{lem1} and Lemma \ref{lem2}, we determine:
	\begin{flalign*}
		&\mathbb{P}(\tau_1>i,\tau_0=i)\approx\frac{1}{4}\sqrt{\frac{\sigma_{i}^{2}}{\sigma_{i}^{2}+2\mu_{i}}}e^{-\frac{\mu_{i}}{4}}\nonumber\\
		&\prod_{j < i, j \in \mathscr{A}} \left\lbrace 
		\begin{array}{l}
			\mathrm{Q}\left(-\sqrt{\frac{\mu_{j}}{2}}\right) - \frac{1}{2} \sqrt{\frac{\sigma_j^2}{\sigma_j^2 + 2\mu_{j}}} e^{-\frac{\mu_{j}^2}{2(\sigma_j^2 + 2\mu_{j})}} \\
			\quad + \frac{1}{4} \sqrt{\frac{\sigma_j^2}{\sigma_j^2 + 2\mu_{j}}} e^{-\frac{\mu_{j}}{4}}
		\end{array}
		\right\rbrace. 
	\end{flalign*}
	
	Combining \cite[Proposition~1]{ref16} with Property~\ref{prop3}, we get: 
	\begin{flalign*}
		\sqrt{\frac{\sigma_{i}^{2}}{\sigma_{i}^{2}+2\mu_{i}}}=&\sqrt{\frac{2^{k_{i}}\sigma_{p}^{2}}{2^{k_{i}}\sigma_{p}^{2}+2\mu_{i}}}\\
		\approx& \sqrt{\frac{2^{k_{i}-1}}{\mu_{i}}}\sigma_{p}-\frac{1}{2}(\frac{2^{k_{i}-1}}{\mu_{i}})^{\frac{3}{2}}\sigma_{p}^{3}+\mathrm{o}(\sigma_{p}^{3}),\\
		e^{-\frac{\mu_{j}^{2}}{2\sigma_{j}^{2}+4\mu_{j}}}=&e^{-\frac{\mu_{j}^{2}}{4\mu_{j}+2^{k_{j}+1}\sigma_{p}^{2}}}\\
		\approx& e^{-\frac{\mu_{j}}{4}}+2^{k_{j}-3}e^{-\frac{\mu_{j}}{4}}\sigma_{p}^{2}+\mathrm{o}(\sigma_{p}^{2}).
	\end{flalign*}
	
	Let $a_i=e^{-\mu_i/4}$, $a_j=e^{-\mu_j/4}$, $b_i=\sqrt{2^{k_i-1}/\mu_i}$, $b_j=\sqrt{2^{k_j-1}/\mu_j}$, $c_j=\mathrm{Q}(-\sqrt{\mu_j/2})$, and $d_j=2^{k_j-3}e^{-\mu_j/4}$. For a given code length $N$ and information set $\mathscr{A}$, we apply a Taylor expansion to each factor and collect terms according to their order in $\sigma_p$. Since the terms involving $d_j$ contribute only to higher-order terms, whose lowest-order term is $\sigma_p^4$, they are absorbed into the remainder term $\mathrm{o}(\sigma_p^3)$. Hence,
	\begin{flalign*}
		D_N(\sigma_p)
		&\approx \sum_{i \in \mathscr{A}} \frac{a_{i} \left( b_{i} \sigma_{p} - \frac{1}{2} b_{i}^{3} \sigma_{p}^{3} + \mathrm{o}(\sigma_{p}^{3}) \right)}{4 \mathrm{BLER_{SC}}} \\
		&\prod_{j < i, j \in \mathscr{A}} \left\lbrace 
		\begin{array}{l}
			c_{j} - \frac{1}{2} \left( b_{j} \sigma_{p} - \frac{1}{2} b_{j}^{3} \sigma_{p}^{3} + \mathrm{o}(\sigma_{p}^{3}) \right) \\
			\times \left(\frac{1}{2}a_{j} + d_{j} \sigma_{p}^{2} + \mathrm{o}(\sigma_{p}^{2}) \right)
		\end{array}
		\right\rbrace\\
		&\approx \sum_{i\in\mathscr{A}} \frac{ A_{i}\sigma_{p} - B_{i} \sigma_{p}^{2} - C_{i} \sigma_{p}^{3} + \mathrm{o}(\sigma_{p}^{3})}{4\mathrm{BLER_{SC}}},
	\end{flalign*}
	where
	\begin{flalign*}
		A_{i} &= a_{i}b_{i}\prod_{j<i,j\in\mathscr{A}}c_{j},\\
		B_{i} &= \frac{1}{4}a_{i}b_{i}\prod_{j<i,j\in\mathscr{A}}a_{j}b_{j},\\
		C_{i} &= \frac{1}{2}a_{i}b_{i}^{3}\prod_{j<i,j\in\mathscr{A}}c_{j}.
	\end{flalign*}
	
	By setting $\frac{\mathrm{d} \mathbb{P}(\mathrm{delay})}{\mathrm{d} \sigma_p} = 0$, we arrive at:
	\begin{flalign*}
		\sum_{i\in\mathscr{A}}A_{i}-2\sum_{i\in\mathscr{A}}B_{i}\sigma_{p}-3\sum_{i\in\mathscr{A}}C_{i}\sigma_{p}^{2}\approx 0.
	\end{flalign*}

	Define $f(\sigma_p)=\sum_{i\in\mathscr{A}}A_i-2\sum_{i\in\mathscr{A}}B_i\sigma_p-3\sum_{i\in\mathscr{A}}C_i\sigma_p^{2}$, and let $\sigma_p^{*}$ denote the positive root of the resulting quadratic approximation. It is straightforward to verify that $f(0)>0$ and $\frac{\mathrm{d}f(\sigma_p)}{\mathrm{d}\sigma_p}<0$ for all $\sigma_p>0$. Hence, $f(\sigma_p)$ is strictly decreasing on $\sigma_p>0$ and changes sign at $\sigma_p=\sigma_p^{*}$. Consequently, $f(\sigma_p)>0$ for $0<\sigma_p<\sigma_p^{*}$ and $f(\sigma_p)<0$ for $\sigma_p>\sigma_p^{*}$.
	
	Since $f(\sigma_p)$ determines the sign of $\frac{\mathrm{d}\mathbb{P}(\mathrm{delay})}{\mathrm{d}\sigma_p}$, it follows that $\mathbb{P}(\mathrm{delay})$ is increasing for $0<\sigma_p<\sigma_p^{*}$ and decreasing for $\sigma_p>\sigma_p^{*}$, which completes the proof of Theorem~\ref{thm1}.
\end{proof}
\noindent\textit{Connection to BLER:}
Adopting $\tau_j=N$ for successful decoding, define
$\mathrm{BLER}_{\mathrm{SC}}:=\Pr(\tau_0<N)$ and
$\mathrm{BLER}_{T=2}:=\Pr(\tau_0<N,\tau_1<N)$ for SC and PE-SC with
one perturbed attempt, respectively. Let
$R_N(\sigma_p):=\Pr(\tau_0<\tau_1<N\mid\tau_0<N)$ denote a delayed FEP
followed by decoding failure. Since a delayed FEP either remains finite
or is eliminated,
\begin{equation}
	\mathrm{BLER}_{T=2}
	=
	\mathrm{BLER}_{\mathrm{SC}}
	\bigl[1-D_N(\sigma_p)+R_N(\sigma_p)\bigr].
	\label{eq:bler_delay}
\end{equation}
Consequently,
$D_N-R_N=\mathbb{P}(\tau_1=N\mid\tau_0<N)$ is the relative BLER reduction,
and $D_N$ is its upper bound.

Theorem~\ref{thm1} reveals that moderate perturbations may delay the
original FEP while preserving useful LLR information, whereas excessive
perturbations may introduce new errors. Additional simulations suggest
that $R_N$ decreases with $N$, making $D_N$ a more accurate BLER
surrogate at larger blocklengths; a rigorous analysis is deferred to an
extended version.

\subsection{Approximate Perturbation-Power Selection}
Theorem~\ref{thm1} yields a closed-form rule for approximately
maximizing the FEP delaying probability, eliminating the need for a
separate selection algorithm. The perturbation power is
selected as
\begin{equation}
	\widehat{\sigma}_{p,\mathrm{op}}^{\,2}
	=
	\left(
	\frac{\sqrt{B_{\Sigma}^{2}+3A_{\Sigma}C_{\Sigma}}-B_{\Sigma}}
	{3C_{\Sigma}}
	\right)^{2},
	\label{eq:power_selection}
\end{equation}
i.e., the square of the positive stationary point in~\eqref{eq11}.
The coefficients are evaluated using the decision-LLR statistics
$\mu_i$ reused from the initial SC attempt and the $g$-function counts
$k_i$, which can be precomputed offline as the Hamming weights of the
binary representations of $i$. Hence, the rule requires only
$\mathscr{O}(K)$ additional arithmetic operations and no additional
decoding traversal, preserving the
$\mathscr{O}(N\log_2N)$ complexity order of SC decoding
\cite[Section~VIII-B]{ref1}.

\section{Simulation and Discussions}
In this section, we present simulation results over a binary-input AWGN channel. The information set $\mathscr{A}$ is constructed using the GA method~\cite{ref14}. An 8-bit CRC with generator polynomial $X^8+X^2+X+1$ is employed for error detection. For each configuration, simulations are terminated after collecting $400$ error packets to ensure reliable performance evaluation~\cite{ref15}.

\begin{figure*}
	\centering
	\includegraphics[width=0.93\textwidth]{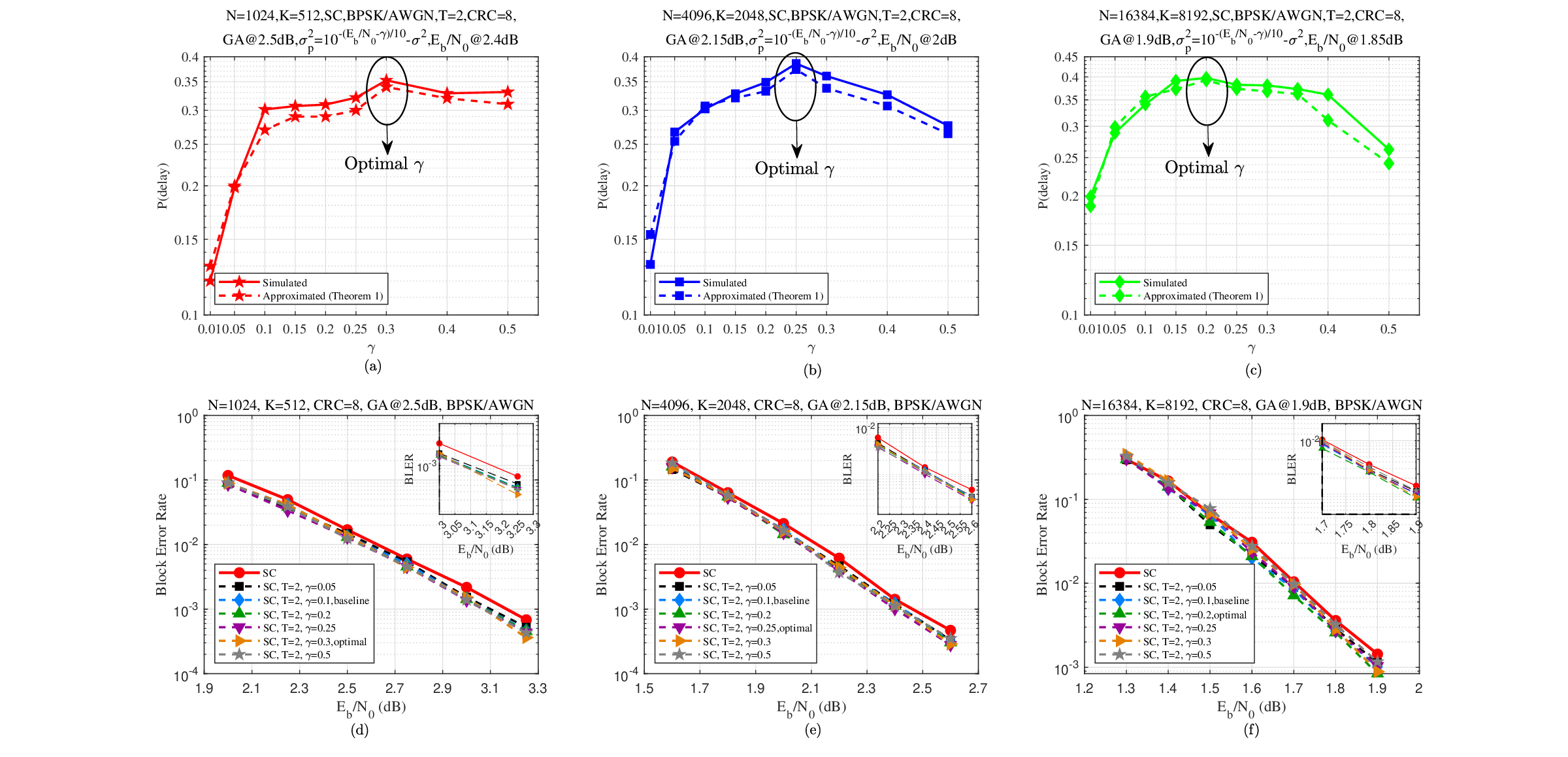}
	\captionsetup{width=0.93\textwidth}
	\caption{Performance evaluation of polar codes. Upper panels (a--c): theoretical optimal perturbation power $\gamma$ (Theorem~1, dashed) versus simulation results (solid). Lower panels (d--f): corresponding BLER versus $\gamma$ for $N\in\{1024,4096,16384\}$.}
	\label{fig2}
\end{figure*}

As shown in Fig.~\ref{fig2}, the theoretical FEP delaying probability
closely matches simulations across all tested settings. Following
\cite[Section~III]{ref14}, the perturbation power is parameterized by
the effective $E_b/N_0$ reduction $\gamma$ as
$\sigma_p^{2}=10^{-(E_b/N_0-\gamma)/10}-\sigma^{2}$.
The small discrepancies mainly arise from the approximation of the
$\mathrm{Q}(\cdot)$-function.

The $\gamma$ maximizing the FEP-delay probability nearly coincides with that minimizing the BLER for all tested block lengths and SNRs. In Fig.~\ref{fig2}, these points are marked by circles in the upper subfigures and by ``optimal'' in the lower ones, while the blue dashed ``baseline'' denotes the setting in~\cite{ref14}. With one perturbed attempt, the proposed selection provides approximately $0.05$--$0.1$ dB gain over conventional SC decoding. Although this agreement is established empirically, it supports the FEP-delay probability as a practical surrogate for BLER-oriented power selection. For multiple perturbed attempts, the same rule can be reused with an independently generated perturbation for each attempt; detailed results are deferred to an extended version due to space limitations.

	\section{Conclusions}
	In this paper, we investigate the impact of perturbation power on the
	FEP delaying probability in PE-SC decoding. We show that properly
	calibrated perturbations improve the FEP-delay probability, whereas
	excessive perturbations degrade decoding performance. Based on a
	third-order approximation, we derive a low-complexity power-selection
	rule using SC decision-LLR statistics and precomputable $g$-function
	counts. We also establish the finite-length relationship between FEP
	delay and BLER reduction. Simulation results show that the selected
	power nearly coincides with the BLER-optimal choice, supporting the
	FEP delaying probability as a practical criterion for finite-length PE-SC
	decoding. Future work will extend the proposed framework to
	multi-perturbation schemes, bit-adaptive strategies, and the design of
	PE-SCL decoding.

	\section*{Acknowledgments}
	This work was supported by the National Key R\&D Program of China under Grant 2023YFA1009601. 
	
%
%

%

		

\newpage
		\begin{thebibliography}{99}
			\bibliographystyle{IEEEtran}
			\balance
			\bibitem{ref1}
			E. Arıkan, “Channel Polarization: A Method for Constructing Capacity-Achieving Codes for Symmetric Binary-Input Memoryless Channels,” \textit{IEEE Trans. Inf. Theory}, vol. 55, no. 7, pp. 3051–3073, Jul. 2009.
			\bibitem{ref2}
			3GPP TSG RAN WG1, R1-167703, “Channel coding scheme for URLLC, mMTC and control channels,” \textit{Intel Corporation}, Aug. 2016. 
			\bibitem{ref3}
			I. Tal, and A. Vardy, “List Decoding of Polar Codes,” \textit{IEEE Trans. Inf. Theory}, vol. 61, no. 5, pp. 2213-2226, May 2015.
			\bibitem{ref4}
			A. Balatsoukas-Stimming, M. B. Parizi, and A. Burg, “LLR-based successive cancellation list decoding of polar codes,” \textit{IEEE Trans. Signal Process.,} vol. 63, no. 19, pp. 5165–5179, Oct. 2015.
			\bibitem{ref5}
			K. Niu, and K. Chen, “CRC-Aided Decoding of Polar Codes,” \textit{IEEE Commun. Lett.}, vol. 16, no. 10, pp. 1668-1671, Oct. 2012.
			\bibitem{adaptive_SCL_CRC}
			B. Li, H. Shen, and D. Tse, “An adaptive successive cancellation list decoder for polar codes with cyclic redundancy check,” \textit{IEEE Commun. Lett.}, vol. 16, no. 12, pp. 2044–2047, Dec. 2012.
			\bibitem{ref6}
			O. Afisiadis, A. Balatsoukas-Stimming, and A. Burg, ``A low-complexity improved successive cancellation decoder for polar codes,'' in \textit{Proc. IEEE Asilomar Conf. Signals Syst. Comput. (Asilomar)}, 2014, pp. 2116–2120.
			\bibitem{ref7}
			L. Chandesris, V. Savin, and D. Declercq, ‘‘Dynamic-SCFlip decoding of polar codes,’’ \textit{IEEE Trans. Commun.}, vol. 66, no. 6, pp. 2333–2345, Jun. 2018.
			\bibitem{T_SCF_Ercan}
			F. Ercan, C. Condo, and W. J. Gross, “Improved bit-flipping algorithm for successive cancellation decoding of polar codes,” \textit{IEEE Trans. Commun.}, vol.67, no. 1, pp. 61–72, Jan. 2019.
			\bibitem{Threshold_SCF}
			Z. Liu, L. Yao, S. Yuan, G. Yan, Z. Ma, and Y. Liu, ``Threshold Successive Cancellation Flip Decoding Algorithm for Polar Codes: Design and Performance,'' \textit{Entropy,} vol. 27, no. 6, pp. 626, June, 2025.
			\bibitem{ref8}
			M. Geiselhart, A. Elkelesh, M. Ebada, S. Cammerer, and S. ten Brink, “On the automorphism group of polar codes,” \textit{in Proc. IEEE Int. Symp. Inf. Theory}, July. 2021, pp. 1230–1235.
			\bibitem{ref9}
			C. Pillet, V. Bioglio and I. Land, “Polar Codes for Automorphism Ensemble Decoding,” \textit{in Proc. IEEE Inf. Theory Workshop (ITW)}, Oct. 2021, pp. 1-6.
			\bibitem{Group_AE_polar}
			V. Bioglio, I. Land, and C. Pillet, “Group properties of polar codes for automorphism ensemble decoding,” \textit{IEEE Trans. Inf. Theory}, vol. 69,	no. 6, pp. 3731–3747, 2023.
			\bibitem{ref10}
			L. Johannsen, C. Kestel, M. Geiselhart, T. Vogt, S. T. Brink, and N. Wehn, “Successive Cancellation Automorphism List Decoding of Polar Codes,” \textit{ in Proc. 12th Int. Symp. Topics Coding (ISTC),}  
			Sep. 2023, pp. 1–5.
			\bibitem{ref11}
			F. Cheng, A. Liu, Y. Zhang, and J. Ren, ‘‘Bit-flip algorithm for successive cancellation list decoder of polar codes,’’ \textit{IEEE Access}, vol. 7, pp. 58346–58352, 2019.
			\bibitem{ref12}
			K. Ivanov and R. Urbanke, “Polar codes do not have many affine automorphisms,” \textit{in Proc. IEEE Int. Symp. Inf. Theory}, Jun. 2022, pp. 2374–2378.
			\bibitem{ref13}
			N. K. Gerrar, S. Zhao, and L. Kong, “CRC-aided perturbed decoding of polar codes,” \textit{in Proc. Int. Conf. Wireless Commun. Netw. Mobile Comput.}, 2018, pp. 18-20.
			\bibitem{ref14}
			X. Wang, H. Zhang, J. Tong, J. Wang, J. Ma, and W. Tong, “Perturbation-enhanced SCL decoder for polar codes,” \textit{in Proc. 2023 IEEE Globecom Workshops (GC Wkshps)}, 
			 Dec. 2023, pp. 1674-1679.
			 \bibitem{ref16}
			 Z. Liu, L. Yao, S. Yuan, G. Yan, Z. Ma and Y. Liu, "Performance Analysis of Perturbation-Enhanced SC Decoders," \textit{IEEE Commun. Lett.}, vol. 29, no. 3, pp. 507-511, Mar. 2025.
			\bibitem{adaptive_PSCL}
			X. Wang, H. Zhang, J. Tong, J. Wang, and W. Tong, “Adaptive perturba
			tion enhanced SCL decoder for polar codes,” 2024, arXiv:2407.03555.
			\bibitem{Yang_Improved_SC_perturb}
			Z. Yang, L. Chen, X. Wang, and H. Zhang, “Improved successive cancellation decoding of long polar codes through perturbing a posteriori LLRs and its theoretical insights,” \textit{IEEE Trans. Commun.}, vol. 74, pp. 5489–5503, 2026.
			\bibitem{ref15}
			P. Trifonov, “Efficient design and decoding of polar codes,” \textit{IEEE Trans. Commun.,} vol. 60, no. 11, pp. 3221–3227, Nov. 2012.
			\bibitem{MSA_f_function_SC}
			C. Leroux, I. Tal, A. Vardy, and W. J. Gross, “Hardware architectures for successive cancellation decoding of polar codes,” \textit{in Proc. IEEE Int. Conf. Acoust., Speech Signal Process.}, May 2011, pp. 1665–1668.
			\bibitem{ref17}
			S. Malluri, and V. K. Pamula, ‘‘Gaussian Q-function and its approximations,’’ \textit{in Proc. Int. Conf. Commun. Syst. Netw. Technol.,} 
			Apr. 2013, pp. 74–77.
			\bibitem{ref18}
			D. Wu, Y. Li, and Y. Sun, “Construction and block error rate analysis of polar codes over AWGN channel based on Gaussian approximation,” \textit{IEEE Commun. Lett.,} vol. 18, no. 7, pp. 1099–1102, Jul. 2014.
		\end{thebibliography}

	\end{document}